\documentclass[twocolumn,showkeys,aps,prb,showpacs]{revtex4-1}
\UseRawInputEncoding
\usepackage{graphicx}
\usepackage[CJKbookmarks,dvipdfm,colorlinks,linkcolor=blue,citecolor=blue]{hyperref}

\begin{document}

\title{Spin-valley-coupled quantum spin Hall insulator with  topological
Rashba-splitting edge states in Janus  monolayer  $\mathrm{CSb_{1.5}Bi_{1.5}}$}

\author{San-Dong Guo}
\affiliation{School of Electronic Engineering, Xi'an University of Posts and Telecommunications, Xi'an 710121, China}

\begin{abstract}
Achieving combination of spin
and valley polarized  states with topological
insulating phase is pregnant to promote the
fantastic integration of topological  physics, spintronics and valleytronics.
In this work, a spin-valley-coupled quantum spin Hall insulator (svc-QSHI) is predicted in Janus  monolayer  $\mathrm{CSb_{1.5}Bi_{1.5}}$ with  dynamic, mechanical and thermal  stabilities.
The inequivalent valleys have opposite Berry curvature and spin
moment, which can produce  a spin-valley Hall effect. In the center of Brillouin zone, a Rashba-type spin splitting can be observed due to missing
horizontal mirror symmetry.
Moreover,  monolayer  $\mathrm{CSb_{1.5}Bi_{1.5}}$ shows unique Rashba-splitting edge
states. Both energy band gap and spin-splitting at the valley point are larger than
the thermal energy of  room temperature (25 meV) with generalized gradient approximation (GGA) level, which is very important  at room temperature for
device applications. It is proved that the spin-valley-coupling and nontrivial quantum spin Hall (QSH) state are robust again  biaxial strain.
Our work may provide a new platform to achieve  integration of topological physics, spintronics and valleytronics.

\end{abstract}
\keywords{Valley, Spin, Topological insulator, Janus structure~~~~~~~~~~~~~~~~~~~~~~~~~Email:sandongyuwang@163.com}

\maketitle

\section{Introduction}
 Since the  valley-dependent effects are discovered
in  $\mathrm{MoS_2}$ monolayer with missing  inversion symmetry,  the field of valleytronics is truly flourishing\cite{q1,q2,q3}.
For hexagonal  two-dimensional (2D)
materials like monolayer $\mathrm{MoS_2}$, the conically shaped valleys at -K
and K corners are inequivalent, and  the spin polarizations  are opposite, as the two points are connected by  the time reversal symmetry operation.
The combination of inversion symmetry breaking and spin-orbit coupling (SOC)
can remove spin degeneracy,  and then gives rise to valley-contrasting
spin splitting, which is the foundation for  spin-valleytronics.
With an applied in-plane electric field, the charge carriers with
opposite valley and spin indexes will attain opposite anomalous transverse velocity, and then a simultaneous spin and
valley Hall effect is produced\cite{q4,q5,q5-1}.

On the other hand, the  topological insulator (TI)  has  spin-momentum-locked conducting
edge states and insulating properties in the bulk, whose  charge and spin transport in the
edge states are quantized dissipationless\cite{t1,t2}. These bring possibilities for low-dissipation
electronic devices.
For 2D materials, the TI is also called as QSH insulator (QSHI)
characterized by counter-propagating edge currents with
opposite spin polarization, which is firstly predicted in graphene\cite{t3}.
Experimentally confirmed QSHIs include  the  HgTe/CdTe and InAs/GaSb quantum wells\cite{t4,t5}, and many other 2D materials
have been proposed as QSHIs by the first-principles calculations\cite{t6,t7,qt3,t8,t9,t10}.
To this end, it's a natural idea to achieve  the integration of QSHI with spin-valleytronics (namely, svc-QSHI).

Several  $\mathrm{AB_3}$ type atomic
sheets have been experimentally synthesized,  for example $\mathrm{BC_3}$ nanosheets\cite{q6}.
Recently, 2D $\mathrm{AB_3}$ monolayers have also  been theoretically
reported\cite{q7,q8,q9,q10}. The   $\mathrm{CP_3}$ monolayer can be used  as anode for sodium-ion batteries\cite{q7},  and the  massless Dirac-Fermions can be achieved in  $\mathrm{CAs_3}$ monolayer\cite{q8}.  In addition to this, the  QSHIs and topological
Rashba-splitting edge states in monolayer
$\mathrm{CX_3}$ (X=Sb and Bi) with inversion symmetry have been predicted\cite{q10}.
The  $\mathrm{MoS_2}$ with 1T' phase is a  QSHI, and the  corresponding Janus structures  MoSSe can still  possess nontrivial topology tuned by strain\cite{gsd4-1}. Compared to  $\mathrm{MoS_2}$, the  MoSSe will lose inversion symmetry.
Inspiring from  this, we construct Janus  monolayer  $\mathrm{CSb_{1.5}Bi_{1.5}}$ based on monolayer
$\mathrm{CX_3}$ (X=Sb and Bi).  By first principles simulations, we
 show that Janus  monolayer  $\mathrm{CSb_{1.5}Bi_{1.5}}$ is a svc-QSHI with topological
Rashba-splitting edge states.  Calculated results show these novel features are robust again  biaxial strain.
These results make monolayer  $\mathrm{CSb_{1.5}Bi_{1.5}}$ an appealing original quantum
material for topological physics, spintronics and valleytronics.

\begin{figure}
  \includegraphics[width=8.0cm]{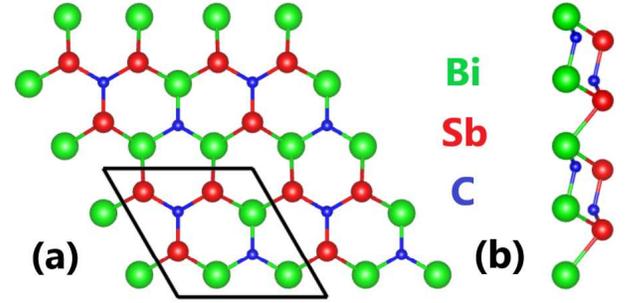}
  \caption{(Color online)The  crystal structure of Janus monolayer  $\mathrm{CSb_{1.5}Bi_{1.5}}$: top view (a) and side view (b).  The  rhombus primitive cell is shown by  black  frames in (a).}\label{t0}
\end{figure}

\begin{figure}
  \includegraphics[width=8cm]{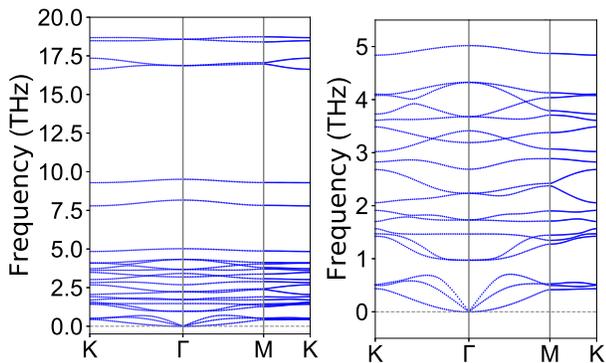}
\caption{(Color online)The  phonon dispersions (Left)  of  Janus monolayer  $\mathrm{CSb_{1.5}Bi_{1.5}}$ using GGA with the enlarged views of low frequency (Right).  }\label{phon}
\end{figure}

\begin{figure}
  \includegraphics[width=8cm]{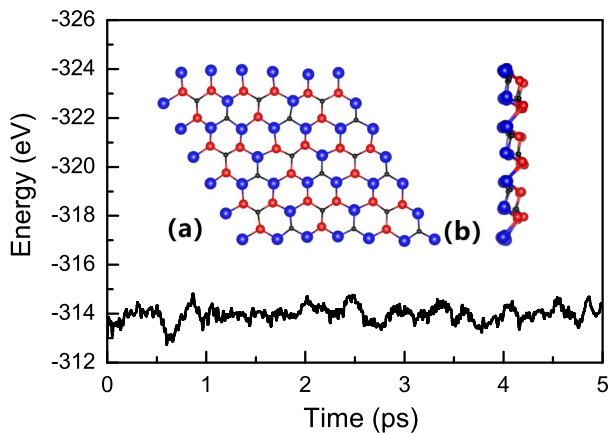}
\caption{(Color online) The variation of free energy during the 5 ps AIMD simulation of Janus monolayer  $\mathrm{CSb_{1.5}Bi_{1.5}}$. Insets show the
 final structures (top view (a) and side view (b)) of $\mathrm{CSb_{1.5}Bi_{1.5}}$ after 5 ps at 300 K. }\label{md}
\end{figure}

\begin{figure}
  \includegraphics[width=8cm]{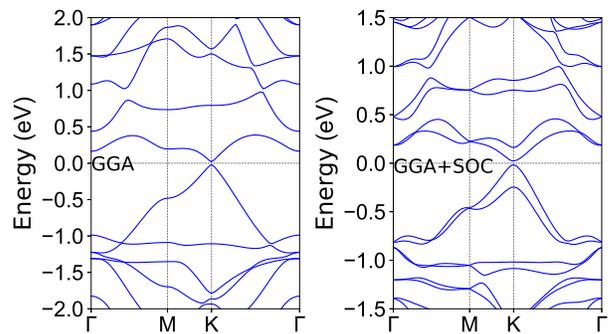}
\caption{(Color online)The energy band structures  of   Janus monolayer  $\mathrm{CSb_{1.5}Bi_{1.5}}$  using GGA  and GGA+SOC.  }\label{band}
\end{figure}

\begin{table}
\centering \caption{For Janus monolayer  $\mathrm{CSb_{1.5}Bi_{1.5}}$, the lattice constants $a_0$ ($\mathrm{{\AA}}$); the structural  parameters including:  C-Sb ($d_1$)  and C-Bi ($d_2$) bond lengths ($\mathrm{{\AA}}$),
Sb-C-Sb ($\theta_1$) and Bi-C-Bi ($\theta_2$) bond angles ($^{\circ}$),  the thickness layer height ($t$) ($\mathrm{{\AA}}$);
 the elastic constants $C_{ij}$ ($\mathrm{Nm^{-1}}$); shear modulus
$G_{2D}$ ($\mathrm{Nm^{-1}}$);  Young's modulus $C_{2D}$  ($\mathrm{Nm^{-1}}$);  Poisson's ratio $\nu_{2D}$. }\label{tab0}
  \begin{tabular*}{0.48\textwidth}{@{\extracolsep{\fill}}cccccc}
  \hline\hline
$a_0$& $d_1$ & $d_2$& $\theta_1$&$\theta_2$&$t$\\\hline
7.772& 2.131 & 2.217 & 115.442 & 115.154 &1.901\\\hline\hline
$C_{11}$/$C_{22}$ &  $C_{12}$& $G_{2D}$&$C_{2D}$& $\nu_{2D}$&\\\hline
26.01&9.13&8.44&22.81&0.351&  \\\hline\hline
\end{tabular*}
\end{table}

\begin{figure*}
  \includegraphics[width=14cm]{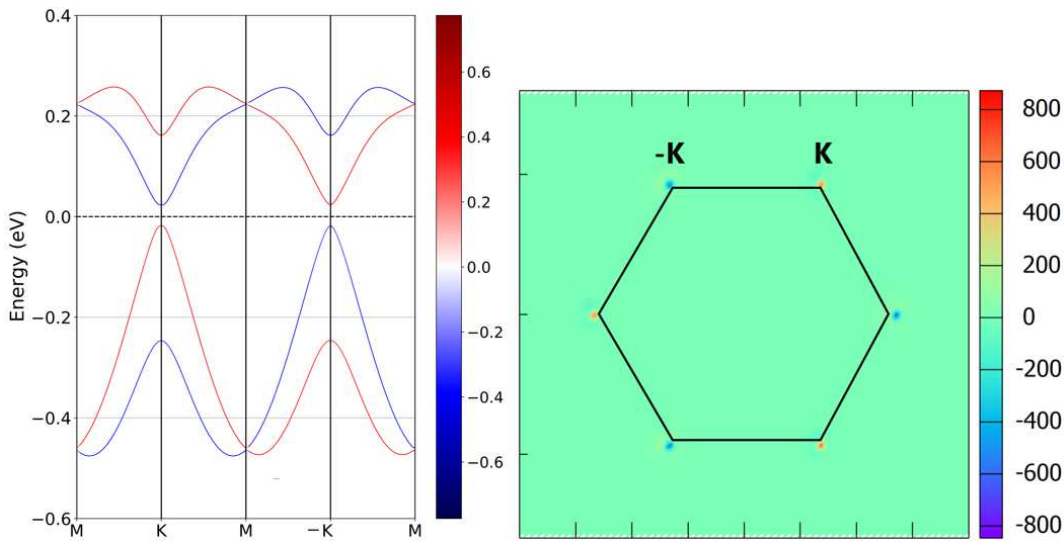}
\caption{(Color online) (Left)The  color map of the band structure of Janus monolayer  $\mathrm{CSb_{1.5}Bi_{1.5}}$ with the
projection of spin operator $\hat{S}_z$, and  the red and blue colors indicate the
spin-up and spin-down bands, respectively. (Right) Calculated Berry curvature distribution of Janus monolayer  $\mathrm{CSb_{1.5}Bi_{1.5}}$  in the 2D Brillouin zone.    }\label{vb}
\end{figure*}

\section{Computational detail}
Within density functional theory (DFT)\cite{1},  we  perform the first-principles calculations  using  the projected augmented wave
(PAW) method, as implemented in
the  VASP package\cite{pv1,pv2,pv3}.   We use  GGA of Perdew, Burke and  Ernzerhof  (GGA-PBE)\cite{pbe}  as the exchange-correlation potential.
 The cutoff energy
for plane-wave expansion is 500 eV  with the total energy  convergence criterion being $10^{-7}$ eV.
To avoid interactions
between two neighboring images, the vacuum
region along the z direction is set to be larger than 18 $\mathrm{{\AA}}$.
The SOC
is incorporated for  band structure
calculations. The Brillouin zone is sampled by using a
12$\times$12$\times$1  K-point meshes  for geometry optimization, elastic coefficients and self-consistent electronic structure calculations.
The geometry optimization is considered
to be converged,  when the residual force on each atom is less than 0.0001 $\mathrm{eV.{\AA}^{-1}}$.
Phonon dispersion spectrum  is attained by the Phonopy code\cite{pv5}  based on  finite displacement method using a 5$\times$5$\times$1 supercell.
The $Z_2$ invariants are used to investigate topological properties of Janus monolayer  $\mathrm{CSb_{1.5}Bi_{1.5}}$,  as implemented  by the Wannier90 and WannierTools
codes\cite{w1,w2}, where a tight-binding Hamiltonian with the
maximally localized Wannier functions is fitted to the first-principles band structures.
We use PYPROCAR code to obtain the constant energy contour plots of the spin
texture\cite{py}.

\section{Crystal and electronic structures}
Based on DFT calculations, the optimized  lattice parameters of Janus monolayer  $\mathrm{CSb_{1.5}Bi_{1.5}}$ is 7.772  $\mathrm{{\AA}}$, which is
between ones of  $\mathrm{CSb_{3}}$ (7.58 $\mathrm{{\AA}}$) and  $\mathrm{CBi_{3}}$ (7.96 $\mathrm{{\AA}}$)\cite{q10}.
As shown  in \autoref{t0}  with top  and side views, each C atom forms three C-Sb or C-Bi
bonds with three neighboring Sb or Bi atoms, and each Sb/Bi atom forms
two Sb-Bi bonds and one C-Sb/Bi bond with neighboring Bi/Sb and
C atoms, respectively.
The symmetry of Janus monolayer  $\mathrm{CSb_{1.5}Bi_{1.5}}$ with space group $P3m1$ (No.156) is lower than that of $\mathrm{CSb_{3}}$/$\mathrm{CBi_{3}}$ monolayer  with space group $P\bar{3}m1$ (No.164) due to
the lack of  space inversion symmetry.
 Monolayer $\mathrm{CSb_{3}}$/$\mathrm{CBi_{3}}$  is
composed of two C atomic layers sandwiched between Sb/Bi atomic layers.
Similar to Janus monolayer MoSSe from $\mathrm{MoS_2}$\cite{e1,e2}, the  Janus monolayer    $\mathrm{CSb_{1.5}Bi_{1.5}}$ can be constructed  by  replacing one of two Sb/Bi  layers with Bi/Sb  atoms in monolayer  $\mathrm{CSb_{3}}$/$\mathrm{CBi_{3}}$.
For Janus monolayer    $\mathrm{CSb_{1.5}Bi_{1.5}}$,
the  inequivalent C-Sb and C-Bi bond lengths (Sb-C-Sb and Bi-C-Bi bond  angles) will be induced due to the difference in atomic sizes and electronegativities of Sb and Bi atoms, and they are
2.131 $\mathrm{{\AA}}$ and 2.217 $\mathrm{{\AA}}$ (115.442$^{\circ}$ and 115.154$^{\circ}$), which gives rise to a built-in electric field.
The symmetry reduce will induce Rashba spin splitting,
valley degree of freedom and  piezoelectric polarizations.

 To confirm the stability of
Janus monolayer    $\mathrm{CSb_{1.5}Bi_{1.5}}$, phonon spectra,  ab initio molecular dynamics (AIMD)
simulations  and elastic constants $C_{ij}$  are carried out.
\autoref{phon} shows that all
phonon branches have no imaginary frequency in the entire Brillouin zone,
suggesting its dynamical stability. It is noted that
two in-plane acoustic branches show linear dispersions, while the ZA  branch corresponding  to the out-of-plane vibrations displays a quadratic dispersion.  These conform to  quadratic dispersion of ZA phonon branch, when  a 2D material  is free of stress\cite{r1,r2}.
The optical branches are well separated from acoustic branches with a gap of 0.29 THz, which prohibits the scattering between acoustic and optical phonon modes. The  vibration of the O atoms   are mainly  at high frequency region.
 As shown in \autoref{md},
after heating at 300 K for 5 ps, neither structure reconstruction nor bond breaking with  small  energy fluctuations for monolayer    $\mathrm{CSb_{1.5}Bi_{1.5}}$ is observed, suggesting
its thermal stability.

 Using Voigt notation, the 2D  elastic tensor with space group $P3m1$  can be reduced into:
\begin{equation}\label{pe1-4}
   C=\left(
    \begin{array}{ccc}
      C_{11} & C_{12} & 0 \\
     C_{12} & C_{11} &0 \\
      0 & 0 & (C_{11}-C_{12})/2 \\
    \end{array}
  \right)
\end{equation}
The  two  independent elastic
constants of  monolayer    $\mathrm{CSb_{1.5}Bi_{1.5}}$ are $C_{11}$=26.01 $\mathrm{Nm^{-1}}$ and $C_{12}$=9.13 $\mathrm{Nm^{-1}}$.
 The shear modulus $G^{2D}$ equals to  $C_{66}$, which can be attained by ($C_{11}$-$C_{12}$)/2, and the corresponding value is  8.44 $\mathrm{Nm^{-1}}$. The calculated $C_{ij}$ satisfy the  Born  criteria of  mechanical stability of a material with hexagonal symmetry\cite{ela}:
 $C_{11}>0$ and  $C_{66}>0$,
  confirming  its mechanical stability.
  The Young's modulus $C_{2D}(\theta)$ can be calculated by the following equation\cite{ela1}:
\begin{equation}\label{c2d}
C_{2D}(\theta)=\frac{C_{11}C_{22}-C_{12}^2}{C_{11}sin^4\theta+Asin^2\theta cos^2\theta+C_{22}cos^4\theta}
\end{equation}
where $A=(C_{11}C_{22}-C_{12}^2)/C_{66}-2C_{12}$.  The  monolayer    $\mathrm{CSb_{1.5}Bi_{1.5}}$ is  mechanically isotropic due to  hexagonal symmetry, and the  $C_{2D}$ is 22.81 $\mathrm{Nm^{-1}}$, which is obviously smaller than those of other 2D materials\cite{y1,y2,y3}, suggesting that monolayer    $\mathrm{CSb_{1.5}Bi_{1.5}}$ is more flexible than other 2D materials.
 The Poisson's ratio $\nu_{2D}(\theta)$ is also isotropic, and can be simply written as:
 \begin{equation}\label{e1}
\nu_{2D}=\frac{C_{12}}{C_{11}}
\end{equation}
 The calculated  $\nu_{2D}$ of monolayer    $\mathrm{CSb_{1.5}Bi_{1.5}}$ is 0.351. The related data are summarized in \autoref{tab0}.
\begin{figure*}
  \includegraphics[width=15cm]{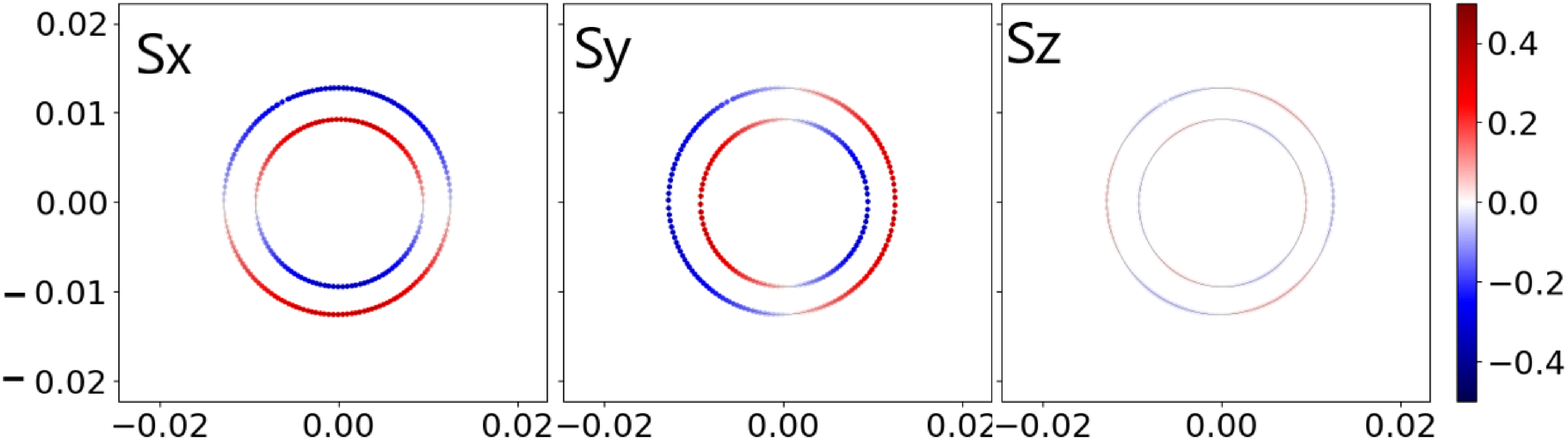}
\caption{(Color online)Spin texture of Janus monolayer  $\mathrm{CSb_{1.5}Bi_{1.5}}$ calculated in the 2D Brillouin zone  centered at the $\Gamma$ point and at an energy surface of 0.25 eV above the Fermi level. The red and blue colours show spin-up and spin-down states, respectively.  }\label{rs}
\end{figure*}

Next, we provide some suggestions on experimental aspects.  Firstly, the $\mathrm{CSb_{3}}$/$\mathrm{CBi_{3}}$ monolayer
may be  experimentally achieved by the bottom-up approaches such as the molecular
beam epitaxy (MBE)\cite{q10}, because
the epitaxial  $\mathrm{BC_3}$ sheet with a similar structure has been successfully synthesized\cite{q6}.
Similar to Janus monolayer MoSSe from $\mathrm{MoS_2}$\cite{e1,e2}, the Janus monolayer  $\mathrm{CSb_{1.5}Bi_{1.5}}$   can be synthesized experimentally with similar experimental techniques based on $\mathrm{CSb_{3}}$/$\mathrm{CBi_{3}}$ monolayer.

\section{Electronic structures}
The energy band structures  of Janus monolayer    $\mathrm{CSb_{1.5}Bi_{1.5}}$ with both GGA and GGA+SOC are plotted in \autoref{band}.
In the absence of SOC, one observes that $\mathrm{CSb_{1.5}Bi_{1.5}}$  is a direct band-gap semiconductor with both
conduction-band minimum (CBM)  and valence-band maximum (VBM) at K point with a gap of 39.9 meV. This is different from that of $\mathrm{CSb_{3}}$/$\mathrm{CBi_{3}}$ with gapless
Dirac points at K points due to symmetry reduce.
Based on  the projected band states to atomic orbitals, the states near the band edges are
dominated by  the C-$p_z$ orbitals  and the Sb/Bi-$p$ orbitals.
When considering SOC, one observes
that $\mathrm{CSb_{1.5}Bi_{1.5}}$ is still a direct band-gap
semiconductor with a gap of 40.1 meV. The main SOC effect  is to give rise to a spin splitting of the
GGA bands.  From the symmetry perspective, the inversion symmetry breaking  lifts the spin degeneracy at each
generic k point. For $\mathrm{CSb_{3}}$/$\mathrm{CBi_{3}}$, no spin degeneracy is removed within SOC  due to existing inversion symmetry\cite{q10}.
The spin-splitting at the K point is as large as 139 meV ($\mathrm{\Delta C}$)  and 229 meV ($\mathrm{\Delta V}$) for the
lowermost conduction band (LCB) and uppermost valence band (UVB), respectively. These splitting energies  are very greater than
the thermal energy of  room temperature (25 meV), which  is highly
desirable for avoiding spin-flip scattering in spintronics applications.

It should be noted that  the conically shaped conduction (valence) band  valleys of Janus monolayer    $\mathrm{CSb_{1.5}Bi_{1.5}}$  at K
and -K corners are inequivalent but related by time-reversal
symmetry.  We redraw the energy band structures with the
projection of spin operator $\hat{S}_z$, including -K high symmetry point (See \autoref{vb}). It is clearly seen  that the spin polarizations at
K and -K are opposite, which  means that the low-energy states
in the K and -K valleys can be distinguished by their spin
index.  Once the K and -K valleys are separated  with a valley polarization,
100\% out-of-plane spin polarization can  be realized in transport. The missing inversion symmetry will make these valleys acquire a valley-contrasting
Berry curvature $\Omega_z(k)$:
\begin{equation}\label{pe1-4}
  \Omega_z(k)=\nabla_k\times i\langle\mu_{n,k}|\nabla_k\mu_{n,k}\rangle
\end{equation}
in which  $\mu_{n,k}$ is the lattice periodic part of the Bloch wave functions.
The distribution of
Berry curvature in the momentum space for monolayer $\mathrm{CSb_{1.5}Bi_{1.5}}$  is plotted in \autoref{vb}.
It is clearly  seen that two obvious peaks at both K and - K valleys but with opposite
sign appear, and  the distribution of $\Omega_z(k)$ exhibits a  three-fold rotational symmetry.   It is not possible to distinguish these two kinds of valleys from energy,  but can discern them by their opposite Berry curvatures and out-of-plane spin moments. This will lead to  spin-valley-coupled transport properties.
When the  in-plane electric field is applied, the valley Hall and spin Hall effects would occur simultaneously due to the valley index being coupled with spin (The charge carriers of different valleys  flow to the opposite
transverse edges due to $\upsilon\sim E\times\Omega_z(k)$), resulting in both valley and spin polarization
along the edges, namely  spin-valley Hall effect.

Moreover, due to the lack of the
horizontal mirror symmetry, Janus monolayer  $\mathrm{CSb_{1.5}Bi_{1.5}}$ should
have Rashba effect. To examine the
Rashba effect,  the in-plane spin-textures are calculated, and the \autoref{rs} shows
the spin projected constant energy (0.25 eV above the Fermi level) contour plots  of the spin textures calculated in $k_x$-$k_y$
plane centered at the $\Gamma$ point.
It is clearly seen that  the pair of spin-splitting bands for both $S_x$ and $S_y$ spin components  have opposite spin orientation.
The pure 2D Rashba spin splitting at the conduction bands around $\Gamma$ point near the Fermi level is observed due to existing concentric spin-texture circles. It is found that   only in-plane $S_x$ and $S_y$ spin components are
present in the Rashba spin split bands with missing out-of-plane $S_z$ component.
 The in-plane spin moments  of two rings have opposite
chirality. The large ring  is  clockwise, while the small ring is anticlockwise.
\begin{figure*}
  \includegraphics[width=14cm]{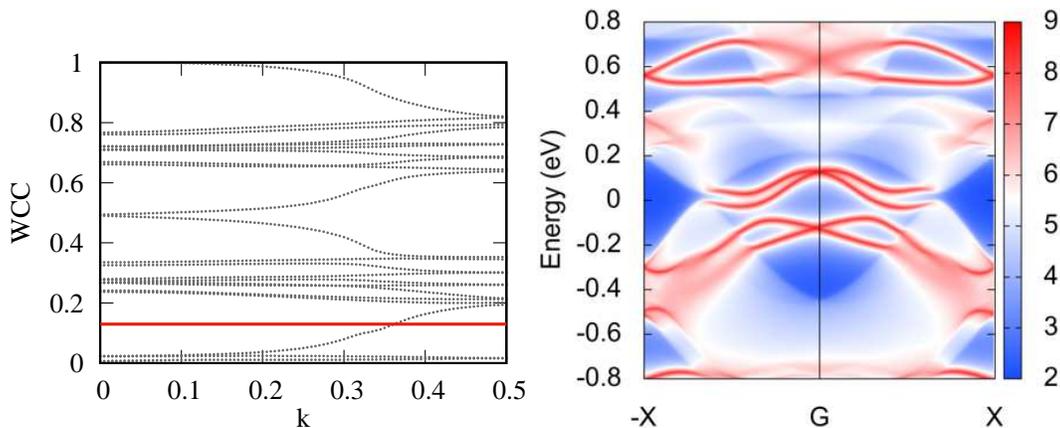}
\caption{(Color online) (Left) Evolution of the WCCs and  (Right)topological edge states connecting the conduction and valence bands of Janus monolayer  $\mathrm{CSb_{1.5}Bi_{1.5}}$.  }\label{zs}
\end{figure*}

The monolayer $\mathrm{CSb_{3}}$/$\mathrm{CBi_{3}}$ is predicted to be a QSHI\cite{q10}.  It is natural to confirm  the topological properties of Janus monolayer  $\mathrm{CSb_{1.5}Bi_{1.5}}$, which  can be characterized by
the $Z_2$ topological invariant. For a material with space inversion symmetry, the $Z_2$ can be calculated by calculating the parities of the occupied valence bands by using  Fu and Kane¡¯s method\cite{f1}, like monolayer $\mathrm{CSb_{3}}$/$\mathrm{CBi_{3}}$\cite{q10}.  The universal method is to
calculate the Wannier Charge Centers (WCCs), which is used for monolayer  $\mathrm{CSb_{1.5}Bi_{1.5}}$.
The evolution
of WCCs for monolayer  $\mathrm{CSb_{1.5}Bi_{1.5}}$ is plotted in \autoref{zs}.  Taking an arbitrary horizontal line  (e.g. WCC=0.13) as reference, one can see that the number of
crossings between the reference line and the evolution of WCCs is odd, which  verifies that $Z_2$=1.  This means that Janus monolayer  $\mathrm{CSb_{1.5}Bi_{1.5}}$  is a QSHI.
Furthermore, a QSHI  should  exhibit  topological protected conducting helical edge states. The Green's-function method is used  to calculate the edge states on (100)  edge based on the tight-binding
Hamiltonian, which are shown in \autoref{zs}. It is clearly seen that
there is a
pair of gapless non-trivial edge states, which  connect the
conduction and valence bands. The edges states  exhibit two quadratic dispersive branches with opposite spin.  The Dirac point is pushed above the Fermi level, and the Rashba-like splitting states can be observed. Similar phenomenon can be found in monolayer $\mathrm{CSb_{3}}$\cite{q10}.

In  practical application,  a substrate  is likely to introduce strain to a 2D material
due to lattice mismatch. A biaxial in-plane strain is  used to study the robustness of the
related physical properties of  Janus monolayer  $\mathrm{CSb_{1.5}Bi_{1.5}}$ against the strain effects.
We use $a/a_0$ (0.94-1.06) to simulate biaxial in-plane strain, where $a$ and $a_0$
represent the in-plane lattice constants  with and
without strain, respectively. The strained energy band structures with both GGA and GGA+SOC are plotted in FIG.1 and FIG.2 of electronic supplementary information (ESI).
Except for 0.94 strain, they all show direct band-gap semiconductors with both
CBM  and VBM at K point. For 0.94 strain, GGA results show a metal, but GGA+SOC results demonstrate an indirect band-gap semiconductor with
CBM  and VBM at $\Gamma$ and  K points.  The energy band gaps with both GGA and GGA+SOC  and spin-splittings at the K point  for LCB and UVB as a function of strain are plotted in FIG.3 and FIG.4 of ESI. These gaps (except for 0.94 strain) and  spin-splittings   are greater than
the thermal energy of  room temperature (25 meV), which is necessary to readily access
and manipulate valleys for memory and logic applications. Finally, we calculate $Z_2$ at all strain points to confirm topological properties  of strained monolayer  $\mathrm{CSb_{1.5}Bi_{1.5}}$, and only show the evolution
of WCCs at 0.94 and 1.06 strains in FIG.5 of ESI. The
calculated results show that the WCCs can be crossed only
one time by an arbitrary horizontal line, which means $Z_2$=1.
These confirm that all strained monolayer  $\mathrm{CSb_{1.5}Bi_{1.5}}$ are TIs.
These  imply  that the spin-valley coupling and  nontrivial topological state are robust against
the biaxial strain.

\section{Discussions and Conclusion}
In summary, our calculated results  demonstrate that  svc-QSHI with spin and valley polarized  states
can emerge in the Janus monolayer  $\mathrm{CSb_{1.5}Bi_{1.5}}$. Particularly,
the $\mathrm{CSb_{1.5}Bi_{1.5}}$ monolayer hosts
Rashba-splitting edges sates, which can be measured by angle-resolved photoemission
spectroscopy (ARPES). Furthermore, we demonstrate that the spin-valley-coupling and
topological properties are perfectly preserved, when a suitable  biaxial strain is applied.
 In view of the recent experimental progress
in  $\mathrm{AB_3}$ type atomic
sheets\cite{q6} and  Janus monolayers\cite{e1,e2}, our findings can promote
further  experimental exploration for intriguing svc-QSHI.

\begin{acknowledgments}
This work is supported by Natural Science Basis Research Plan in Shaanxi Province of China  (2021JM-456). We are grateful to the Advanced Analysis and Computation Center of China University of Mining and Technology (CUMT) for the award of CPU hours and WIEN2k/VASP software to accomplish this work.
\end{acknowledgments}

\end{document}